\documentclass[prl,twocolumn,nofootinbib,showpacs,superscriptaddress,amsmath,amssymb,amsfonts]{revtex4}

\usepackage{ulem}
\usepackage{bm}
\usepackage{color,xcolor}
\usepackage{mathtools,amsbsy}
\usepackage{keyval,graphicx}
\usepackage{slashed}
\usepackage{epstopdf}
\usepackage{isotope}
\usepackage{dcolumn}

\newcommand{\X}{X(3872)}

\newcommand{\Zc}{Z_c(3900)}

\newcommand{\be}{\begin{equation}}
\newcommand{\ee}{\end{equation}}
\newcommand{\ba}{\begin{array}{cc}}
 \newcommand{\ea}{\end{array}}
\newcommand{\bea}{\begin{eqnarray}}
  \newcommand{\eea}{\end{eqnarray}}

\begin{document}
\title{Isospin analysis of $B\to D^*\bar{D}K$ and the absence of the $\Zc$ in $B$ decays}

\author{Zhi Yang}\email{zhiyang@hiskp.uni-bonn.de}
\affiliation{Helmholtz-Institut f\"ur Strahlen- und Kernphysik and Bethe
Center for Theoretical Physics, \\Universit\"at Bonn,  D-53115 Bonn, Germany}
\affiliation{Institute of Modern Physics, Chinese Academy of Sciences, Lanzhou 730000, China}
\author{Qian Wang}\email{wangqian@hiskp.uni-bonn.de}
\affiliation{Helmholtz-Institut f\"ur Strahlen- und Kernphysik and Bethe
Center for Theoretical Physics, \\Universit\"at Bonn,  D-53115 Bonn, Germany}
\author{Ulf-G.~Mei{\ss}ner}\email{meissner@hiskp.uni-bonn.de}
\affiliation{Helmholtz-Institut f\"ur Strahlen- und Kernphysik and Bethe
Center for Theoretical Physics, \\Universit\"at Bonn,  D-53115 Bonn, Germany}
\affiliation{Institut f\"{u}r Kernphysik, Institute for Advanced
Simulation, and J\"ulich Center for Hadron Physics,\\
Forschungszentrum J\"ulich,  D-52425 J\"{u}lich, Germany}

\begin{abstract}

We study the isospin amplitudes in the exclusive $B$ to $D^*\bar{D}K$ decay process and fit the 
available $D^{*0}\bar{D}^0$ invariant mass distributions near threshold. The analysis
demonstrates that the production of the isospin triplet $D^*\bar{D}$ state is highly suppressed
compared to the isospin singlet one. 
That explains why the $\Zc$ has not been found in $B$ decays.
In addition, the production of the negative charge-parity state might be further suppressed 
in the heavy quark limit. These two reasons which are based on the molecular assumption
offer the first explanation why the $\Zc$ is absent in $B$ decays. Further studies of the
absence from both the experimental and the theoretical side is extremely important for understanding
the nature of the $\X$ and the $\Zc$.

\end{abstract}

\date{\today}
\maketitle


In 2013, the BESIII and Belle Collaborations reported the charged
charmonium-like state $\Zc^{\pm}$ in the $\pi^{\pm}J/\psi$ invariant mass distribution 
of the $e^+e^-\to\pi^+\pi^-J/\psi$ process~\cite{Ablikim:2013mio,Liu:2013dau}. 
The observed channel, i.e. $\pi^\pm J/\psi$, reveals its minimal 
four-quark $c\bar{c}u\bar{d}$ constituent nature,
making it more intriguing than other exotic candidates.  
The charged state was also confirmed by the reanalysis of the CLEOc data~\cite{Xiao:2013iha}, 
which also discovered its neutral partner.  
In addition to the $\pi J/\psi$ invariant mass distribution in the 
$e^+e^-\to \pi\pi J/\psi$ process,  the $\Zc$ was also observed in the $D^*\bar{D}$ 
channel\footnote{Here and in what follows, the
charge-conjugated channels are considered implicitly.} by the BESIII
Collaboration~\cite{Ablikim:2015gda,Ablikim:2013xfr}.
The angular analysis of Refs.~\cite{Ablikim:2013xfr,Ablikim:2015swa} leads to its
quantum numbers $I^G(J^P)=1^+(1^+)$. 
The averaged mass is $3886.6\pm2.4~\mathrm{MeV}$~\cite{Olive:2016xmw}
which is slightly above the $D^*\bar{D}$ threshold, thus it is naturally to be regarded as
a $D^*\bar{D}$ molecule 
state~\cite{Wang:2013cya,Guo:2013sya,Wilbring:2013cha,Dong:2013iqa,Zhang:2013aoa,Ke:2013gia}.
This proximity to the threshold also allows for an interpretations as a cusp 
effect~\cite{Swanson:2014tra,Chen:2013coa}.
However, it has been demonstrated that the treatment within the cusp scenario is 
not self-consistent~\cite{Guo:2014iya}
and the near threshold pronounced structure in elastic channels necessary requires a nearby pole. 
Besides these two interpretations, there are also others, such as 
tetraquark~\cite{Faccini:2013lda,Dias:2013xfa,Qiao:2013raa,Wang:2013vex,Deng:2014gqa},
hadro-charmonium~\cite{Voloshin:2013dpa}, and hybrid~\cite{Braaten:2013boa}.

In the $D^*\bar{D}$ hadronic molecular picture, 
the isospin triplet $\Zc$ and the isospin singlet~\cite{Aubert:2004zr,Choi:2011fc} $\X$
share similar dynamics, such as the analogy of  the processes $e^+e^-\to \gamma\X$ and
  $e^+e^-\to \pi \Zc$~\cite{Guo:2013nza, Ablikim:2013dyn}. 
 As a conclusion, both the $\Zc$ and the $\X$ have been observed in $e^+e^-$ annihilation. 
The most puzzling aspect of the $\Zc$ is its absence in $B$ decays which is different
 to the case of the $X(3872)$.  The $\Zc$ is expected to be seen in
the decay $B\to KJ/\psi\pi$ due to the analogy of the $\Zc$ and the $\X$, 
since the latter one has been observed in both $B\to K\X$~\cite{Choi:2003ue}
and $e^+e^-\to\gamma \X$~\cite{Ablikim:2013dyn} processes through
its decay $X(3872)\to\pi^+\pi^-J/\psi$.

\begin{figure}
\begin{center}
  \includegraphics[width=0.515\textwidth]{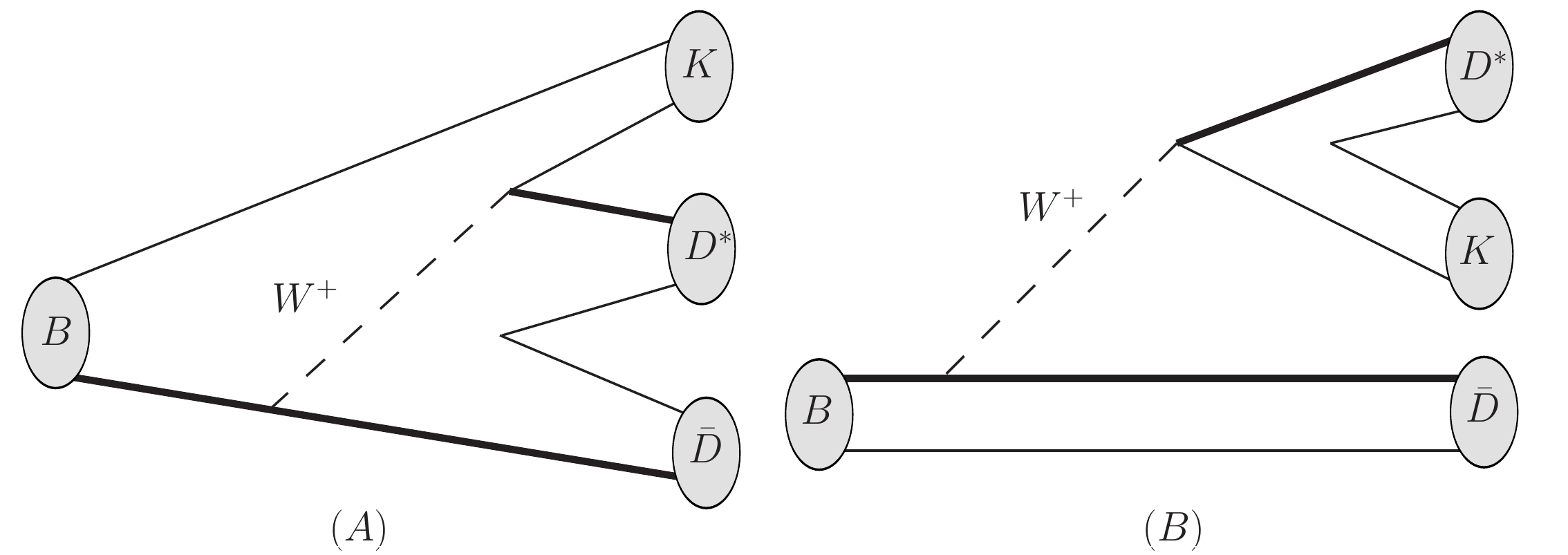}
\caption{Feynman diagrams for B decay to $D^*\bar{D}K$. Diagram (A) is the 
one with an internal $W$-emission,
while diagram (B) corresponds to the one with an external $W$-emission. 
The thick solid lines denote a heavy $b$ or $c$ quark. 
The thin solid and dashed lines represent the light quarks and 
$W$ bosons, respectively.} \label{fig:Bdecay1}
\end{center}
\end{figure}

In the molecular picture, the $\X$ and the $\Zc$ have the same 
constituents $D^*$ and $\bar{D}$, but different isospins and C-parities. Therefore, 
both of them can be formed from the interaction between $D^*$ and
$\bar{D}$. The production of the $\X$ in the process $B\to D^*\bar{D}K$ has already
been studied in Ref.~\cite{Braaten:2004fk}, where it 
 occurs through the
isospin conserved weak transition ($\Delta I=0$) $b\to c\bar{c}s$ current. 
On the quark level, the $\Delta I=0$ process is given in terms of two diagrams,
i.e. the color suppressed internal $W$-emission and the external $W$-emission diagram,  
cf. Fig.~\ref{fig:Bdecay1}.  
Besides those two,  the $\Delta I=1$ diagram, i.e. the second diagram of Fig.~2 
in Ref.~\cite{Zito:2004kz}, could also contribute. 
However, it is CKM suppressed diagram and can safely be neglected.
Further, the isospin decomposition of the penguin diagram, i.e. the first 
diagram of Fig.~2 in Ref.~\cite{Zito:2004kz}, 
is the same as diagram (B) in Fig.~\ref{fig:Bdecay1}. 
Hence the contribution of the penguin diagram in the threshold region
can be absorbed  into the parameters of diagram (B) in Fig.~\ref{fig:Bdecay1}. 
Based on the above analysis, the $B\to D^*\bar{D}K$ process conserves 
isospin ($\Delta I=0$) to the leading order of
 the expansion parameter of the Wolfenstein parametrisation of
  the CKM matrix, although it is a weak decay process.
It is the prerequisite that one can analyse the isospin amplitudes of 
the $B\to D^*\bar{D}K$  process.

First, we derive the isospin relations for the $B\to D^*\bar{D}K$ process through 
the quark-level Feynman diagrams, c.f. Fig.~\ref{fig:Bdecay1}. 
Since the light quark pair created from the vacuum is a flavor and isospin singlet, 
the decay amplitudes of the $B^0$ meson from the two diagrams of Fig.~\ref{fig:Bdecay1} are
\bea
{\cal M}[B^0\to D^{*0} {D}^{-}K^+]&=&-\frac{1}{\sqrt{2}}B_1 ,\label{eq:1}\\
{\cal M}[B^0\to D^{*+} {D}^{-}K^0]&=&\frac{1}{\sqrt{2}}A_0+\frac{1}{2}(B_0+B_1)e^{i\theta},\\
{\cal M}[B^0\to D^{*0} \bar{D}^{0}K^0]&=&-\frac{1}{\sqrt{2}}A_0,
\eea
where $B_0$ ($B_1$) is the amplitude producing the $D^*\bar{D}$ with isospin $0$ ($1$) through
external $W$-emission, $A_0$ corresponds to the internal $W$-emission with isospin 0, 
and $\theta$ is the relative phase between diagram~(A) and diagram~(B) in Fig.~\ref{fig:Bdecay1}. 
The decay amplitudes of the $B^+$ are
\bea
{\cal M}[B^+\to D^{*+} \bar{D}^{0}K^0]&=&\frac{1}{\sqrt{2}}B_1,\label{eq:4}\\
{\cal M}[B^+\to D^{*0} \bar{D}^{0}K^+]&=&\frac{-1}{\sqrt{2}}A_0+\frac{1}{2}(B_0-B_1)e^{i\theta},\\
{\cal M}[B^+\to D^{*+} {D}^{-}K^+]&=&\frac{1}{\sqrt{2}}A_0.
\eea
Here, the isospin relations we derived are different from those in Ref.~\cite{Zito:2004kz}, 
where  the fact that the $D^*\bar{D}$ system from the internal $W$-emission 
can only be isospin $I=0$ due to the light quark pair coming from vacuum was ignored.

\begin{figure}
\begin{center}
  \includegraphics[width=0.48\textwidth]{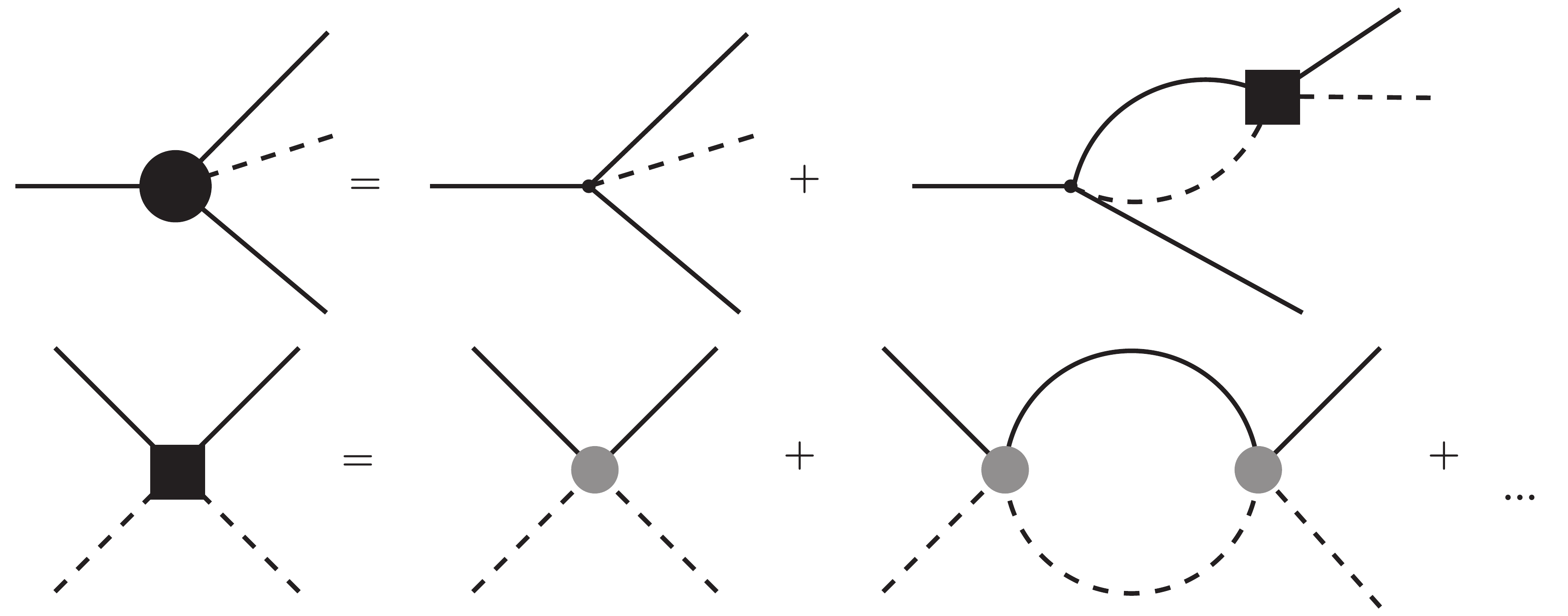}
\caption{Feynman diagrams for the production of $D^*\bar{D}$ through $B$
decay, including the $D^*\bar{D}$ final state interaction, where the dots and the boxes represent the short-distance
and the resummed $D^*\bar{D}$ scattering amplitudes (first row), respectively. The 
resummed $D^*\bar{D}$ scattering amplitude with blobs representing the $D^*\bar{D}\to D^*\bar{D}$ 
vertex is shown in the second row.} \label{fig:Bdecay2}
\end{center}
\end{figure}

Second, except for the short-distance direct production, final-state interactions 
also contribute as shown in Fig.~\ref{fig:Bdecay2}.  The amplitudes consist of 
two parts: one is the $D^*\bar{D}$ short-distance production 
amplitude and the other one is the long-distance $D^*\bar{D}$ scattering. 
The rescattering process proceeds in
two pathways. Taking $D^{*0}\bar{D}^0$ production as an example, it includes
the $D^{*0}\bar{D}^0\to D^{*0}\bar{D}^0$ scattering
following the $B\to D^{*0}\bar{D}^0K$ decay  and $D^{*+}D^-\to D^{*0}\bar{D}^0$ scattering following 
the $B\to D^{*+}D^-K$ decay.
The $D^*\bar{D}$ system we are concerned with is in the near-threshold energy region, 
as we  consider the coalescence of the charm mesons into $\X$ or $\Zc$. At 
the $D^*\bar{D}$ threshold,
Lorentz invariance requires the short-distance decay amplitude to have the
simple form~\cite{Braaten:2004fk}
\be
\begin{split}
{\cal A}_{\text{short}}[B^{0(+)}\to D^{*0}\bar{D}^{0}K^{0(+)}]=c^0_{0(+)} P\cdot \epsilon^*,\\
 {\cal A}_{\text{short}}[B^{0(+)}\to D^{*+}{D}^{-}K^{0(+)}]=c^c_{0(+)} P\cdot \epsilon^*,
\end{split}
\label{eq:Bdecay}
\ee
where $\epsilon$ is the polarization four-vector of the vector charmed meson, 
$P$ is the momentum of the bottom meson and $c^{0(c)}_{0(+)}$ are  coefficients that
need to be determined\footnote{One should notice that the constants are the normalized 
ones as discussed in Refs.~\cite{Braaten:2004fk,Braaten:2004ai}.}.
$c^{0(c)}_{0(+)}$ are the combinations of $a_0$, $b_0$ and $b_1$, which are the constants
corresponding to $A_0=a_0P\cdot \epsilon^*$ and $B_{0(1)}=b_{0(1)}P\cdot \epsilon^*$.

After integrating over phase space, the differential decay width is written as
\be
\frac{d\Gamma}{d M}=\frac{\mu \lambda^{3/2}(m_B,M,m_K)}
{256\pi^3 m_B^3 M^2}\lambda^{1/2}(M,m_{D^{*0}},m_{\bar{D}^{0}})|{\cal A}(E)|^2,
\label{eq:partialwidth}
\ee
where $M$ is the invariant mass of  $D^{*0}$ and $\bar{D}^0$, $E$ is the energy of
the $D^{*0}\bar{D}^0$ system in its rest frame relative to the $D^{*0}\bar{D}^0$ 
threshold, $E=M-(m_{D^{*0}}+m_{\bar{D}^{0}})$,
$\lambda(x,y,z)=x^4+y^4+z^4-2(x^2 y^2
+y^2 z^2+z^2 x^2)$ is the triangle function and ${\cal A}(E)=c^0_{+}+c^0_{+}
{\cal T}_{1,1}+c^c_+ {\cal T}_{2,1}$ for $B^{+}\to D^{*0}\bar{D}^{0}K^{+}$, while
${\cal A}(E)=c^0_{0}+c^0_{0}{\cal T}_{1,1}+c^c_{0}{\cal T}_{2,1}$ for $B^{0}\to D^{*0}\bar{D}^{0}K^{0}$.
The matrix ${\cal T}(E)$ is the two-body scattering amplitude for the 
coupled channels $D^{*0}\bar{D}^0$ and $D^{*+}D^-$~\cite{Cohen:2004kf,Braaten:2005jj}
\be
\frac{1}{{\cal T}(E)}=\frac{1}{2\pi}
\left(
\ba
\mu_1 (-1/a_{11}-i p_0) & \sqrt{\mu_1 \mu_2}/a_{12} \\
\sqrt{\mu_1 \mu_2}/a_{12} & \mu_2 (-1/a_{22}-i p_c)\\
\ea
\right)
\ee
where $p_0=\sqrt{2\mu_1 E}$ and $p_c=\sqrt{2\mu_2 (E-\Delta)}$ are the binding momenta for
the neutral $D^{*0}\bar{D}^0$ and charged $D^{*+}D^-$ channels, respectively, and
 $\Delta=m_{D^+}+m_{D^{*-}}-m_{D^0}-m_{D^{*0}}$ is the energy gap between the two channels.
 Thus, one can fit the $D^*\bar{D}$ invariant mass distributions through Eq.~\eqref{eq:partialwidth}.

Next we turn to the production of the $D^*\bar{D}$ hadronic molecule $\X$. Its production
can be factorized as the short-distance production of the constituents and the 
long-distance formation of the $\X$ state parts. The factorization formulas for the 
prompt production of the
$\X$ at hadron colliders have been studied in Refs.~\cite{Artoisenet:2009wk,Guo:2014sca}.
For the production of $\X$ through $B^0$ or $B^+$ decay, the factorization formula is written as
\be
\Gamma^{0(+)}=\frac{\lambda^{3/2}(m_B,m_X,m_K)}{32\pi m^3_B m_X}\big|c^0_{0(+)}g_0
+c^c_{0(+)}g_c\big |^2,
\ee
where $g_0$ and $g_c$ are the coupling constants of the $D^{*0}\bar{D}^0$ and $D^{*+}D^-$ 
channel to the $\X$ state, respectively, which are related to the residue of the scattering 
amplitudes at the pole position.
The ratio of the $\X$ production through $B^0$ and $B^+$ decays~\cite{Choi:2011fc}
\be
\frac{{\cal B}(B^0\to XK^0)}{{\cal B}(B^+\to XK^+)}=0.50\pm0.14\pm0.04
\ee
 is  another constraint in the fit.

\begin{figure}[tb]
\begin{center}
  \includegraphics[width=0.4\textwidth]{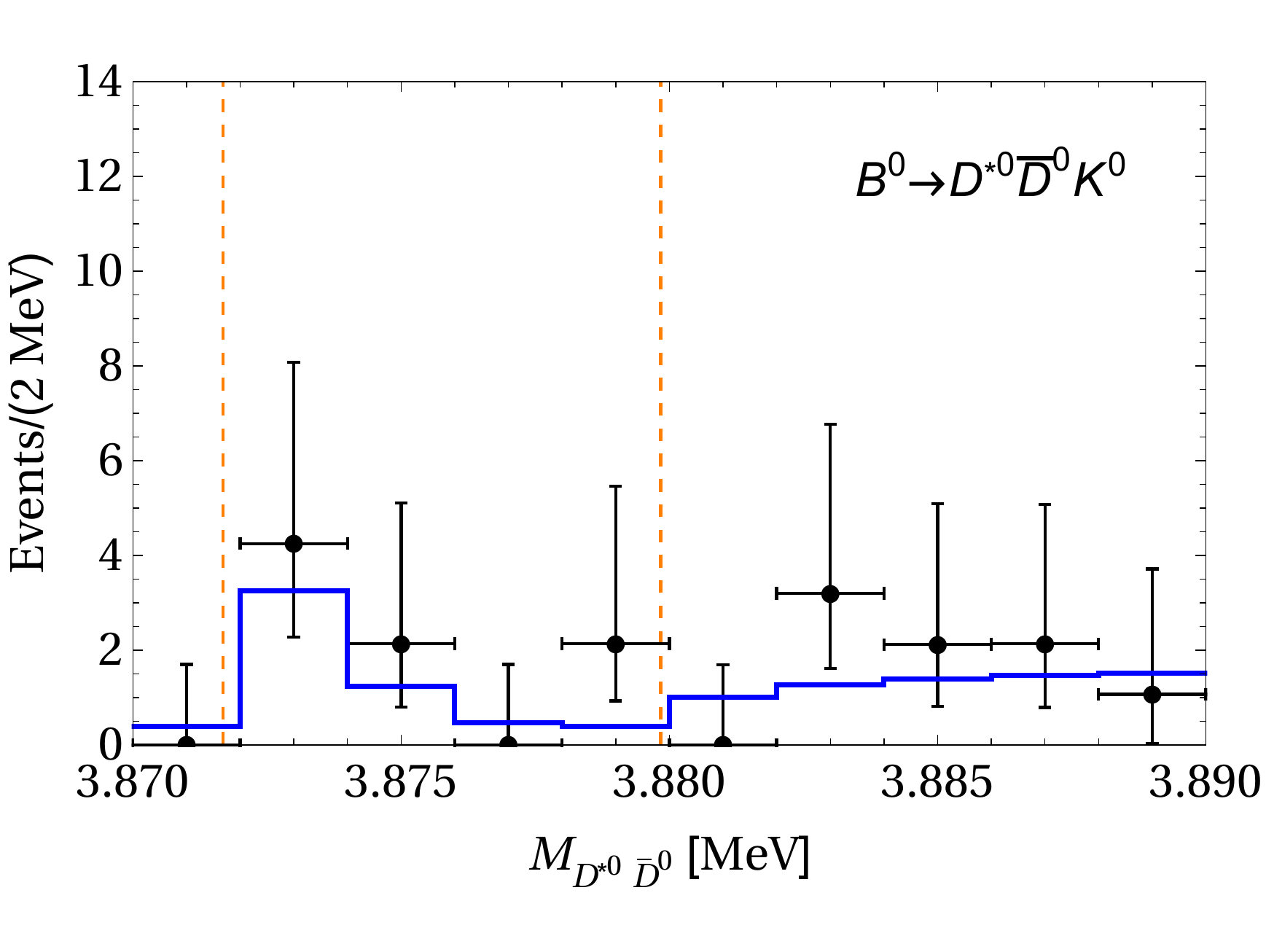}
    \includegraphics[width=0.4\textwidth]{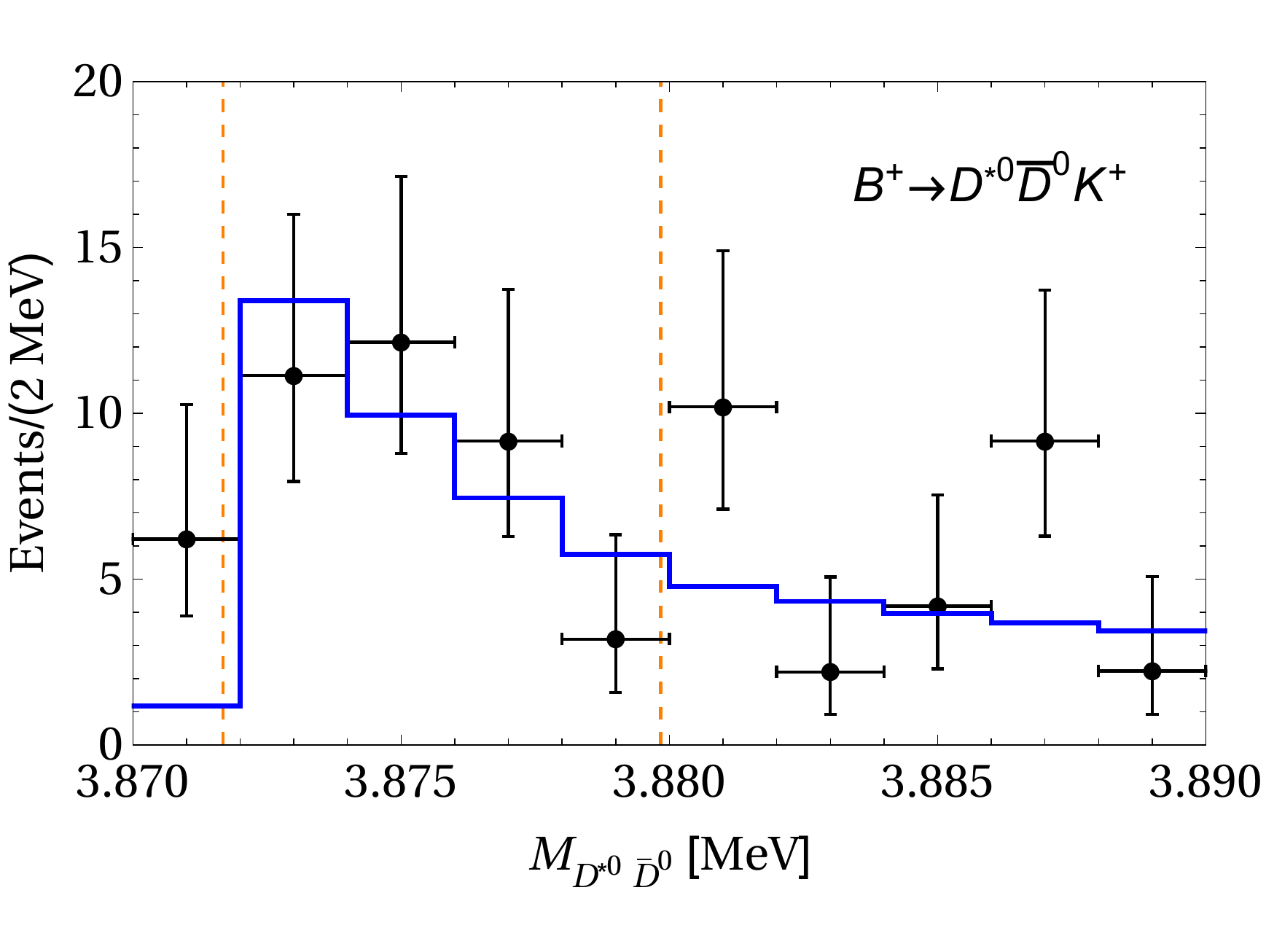}
\caption{Invariant mass distributions of $D^{*0}\bar{D}^0$ for $B^0\to D^{*0}\bar{D}^0K^0$ 
 (up) and $B^+\to D^{*0}\bar{D}^0K^+$ (down) decays. The points with error bars
are the experimental data taken from Ref.~\cite{Zwahlen:2008zz} and the histogram is the fit results. The two vertical
dashed lines are the $D^{*0}\bar{D}^0$ and $D^{*+}D^-$ thresholds. 
Here we use the events of the $D^{*0}\bar{D}^0$ distributions from both $D^{*0}\to D^{0}\gamma$ and 
$D^{*0}\to D^{0}\pi^0$ channels} \label{fig:diff}
\end{center}
\end{figure}

\begin{table}[t]
\renewcommand{\arraystretch}{1.3}
\begin{tabular}{|l|l|}
\hline
Parameter  & value  \\
\hline
$|a_0|$ &  $(2.23\pm 1.02) N~\mathrm{GeV^{-1}}$ \\
$\theta$ &  $0.46\pm 0.43~\mathrm{rad}$\\
$|b_0|$ & $(5.00\pm 1.10) N~\mathrm{GeV^{-1}}$\\
$|b_1|$ & $0.014^{+3.84}_{-0.014} N~\mathrm{GeV^{-1}}$\\
$a_{11}$ & $-1.56 \times 10^{11} \pm 0.28 ~\mathrm{fm}$\\
$a_{12}$ & $3.37\pm 0.27~\mathrm{fm}$\\
$a_{22}$ & $0.94\pm 0.04~\mathrm{fm}$\\
\hline
$\chi^2/n_{d.o.f.}$ & 7.44/14\\
\hline
\end{tabular}
\caption{The fit parameters, where $N$ is an unknown factor related to the efficiency in
the experiment, $a_0$ and $b_0$ ($b_1$) are the production strengths of diagram (A) and (B)
with $I=0$ ($I=1$).  In the fit, the coefficients $a_0$ and $b_{0(1)}$ are written in
polar coordinates with the arguments absorbed
into the relative angular variable $\theta$.}
\label{tab:fitresult}
\end{table}

\begin{table}
\begin{tabular}{|c|ccccccc|}
\hline 
 & $|a_{0}|$ & $\theta$ & $|b_{0}|$ & $|b_{1}|$ & $a_{11}$ & $a_{12}$ & $a_{22}$\tabularnewline
\hline 
$|a_{0}|$ & 1.0 & -0.47 & 0.81 & 0.03 & 0.00 & -0.09 & 0.06\tabularnewline 
$\theta$ &  & 1.0 & -0.80 & -0.05 & 0.00 & 0.07 & -0.06\tabularnewline 
$|b_{0}|$ &  &  & 1.0 & -0.02 & 0.00 & 0.02 & 0.02\tabularnewline 
$|b_{1}|$ &  &  &  & 1.0 & 0.00 & 0.02 & 0.02\tabularnewline
$a_{11}$ &  &  &  &  & 1.0 & 0.00 & 0.00\tabularnewline
$a_{12}$ &  &  &  &  &  & 1.0 & 0.74\tabularnewline
$a_{22}$ &  &  &  &  &  &  & 1.0\tabularnewline
\hline 
\end{tabular}
\caption{Parameter correlation matrix. As the matrix is symmetric,  
only the matrix elements in the upper triangle are presented explicitly. }
\label{tab:fitcorrelation}
\end{table}


The fitted invariant mass distributions are shown as the blue histograms in Fig.~\ref{fig:diff}, 
which pass through all the experimental data. The dip structure at $3.88~\mathrm{GeV}$ in the
$B^0\to D^{*0}\bar{D}^0K^0$ process is due the opening of the charged $D^{*+}D^-$ threshold.
The fit parameters with $1\sigma$ error are shown in Tab.~\ref{tab:fitresult}.
The large uncertainties stem from those of the experimental data. 
The corresponding parameter correlation matrix is shown in Tab.~\ref{tab:fitcorrelation},
from which one can see that there are no particularly large correlations between the parameters. 
That also indicates there are no redundant parameters in our formulae.
Further high statistics data  will help to reduce the uncertainties.
The most interesting parameters are $a_0$, $b_0$ and $b_1$ with the isospin of
the $D^*\bar{D}$ system as subscripts,
which reflect the production strengths of diagrams (A) and (B). 
From the fiited parameters, one can extract several interesting features which help us 
understand the nature of the $\X$ and the absence of the $\Zc$ in $B$ decay.
  \begin{itemize}
  \item The $\X$ behaves as a bound state with the binding energy 
  $\epsilon=m_{D^{*0}}+m_{D^0}-E_\text{pole}=1.06_{-0.50}^{+19.03}~\mathrm{MeV}$
   below the $D^{*0}\bar{D}^0$ threshold. It should not be the same as 
   that, i.e. Eq.~(53) of Ref.~\cite{Guo:2017jvc}, extracted from the mass (or the peak position in another word) of the $\X$.
    That is because that the peak positions are channel dependent, but the pole position is not. 
   Especially, in its component channel, 
   the peak position might be a little higher than the pole position due to the phase space limit.
   \item The contribution of diagram~(B) 
    to the isospin triplet channel is about three orders 
   smaller than that to the isospin singlet channel, as
   $b_1/b_0\sim \mathcal{O}(10^{-3})$. 
   \item 
 As shown by Eqs.~\eqref{eq:1} and \eqref{eq:4}, the small value of $b_1$ also indicates the 
   small short-distance production 
   amplitudes of the $B^0\to D^{*0}D^-K^+$ and $B^+\to D^{*+}\bar{D}^0 K^0$ processes at threshold.
   Thus the production of isotriplet state through these processes is quite small.
      \item Since diagram (A) also contributes to the isospin singlet channel, 
   there is the coherent interference between diagrams (A) and (B) for the isospin singlet channel. 
   As the result, for each individual channel, the ratios of the $I=1$ and $I=0$ components are
\bea\nonumber
B^0\to D^{*+} {D}^{-}K^0: &&\bigskip \\\nonumber
 \frac{|B_1/2|^2}{|A_0/\sqrt{2}+B_0 e^{i\theta}/2|^2}&=&(3.30\times 10^{-6})^{+0.30}_{-3.30\times 10^{-6}},\\\nonumber
B^+\to D^{*0} \bar{D}^{0}K^+: &&\bigskip\\\nonumber
 \frac{|B_1/2|^2}{|-A_0/\sqrt{2}+B_0 e^{i\theta}/2|^2}&=&(3.11\times 10^{-5})^{+1.24\times 10^3}_{-3.11\times 10^{-5}}.
\eea
From the above equations, one concludes that the production of $I=1$
$D^{*0}\bar{D}^0$ and $D^{*+}D^-$ pairs in $B$ decay is highly suppressed. 
The large uncertainty of the upper limit in the second ratio stems 
from the destructive interference in the denominator, as the  errors of the parameters are almost the same, thus
generating this uncertainty. 
  \end{itemize}
 
Besides the isospin suppression, there might be another suppression coming from the C-parity
as discussed in Ref.~\cite{Braaten:2004fk}, where the productions of the states with $C=+$ and $C=-$ are constructive and
destructive, respectively.  In the heavy quark limit,
the wave functions of $D$ and $D^*$ are the same, leaving the production of a state with $C=-$
in the transition from $B$ to $K$ equal to zero as shown by Eq.(8) in Ref.~\cite{Braaten:2004fk}.
However, the branching ratio of $B^+\to \bar{D}^0 D^{*0}K^+$ is 
about 3 times as that of $B^+\to \bar{D}^{*0}D^0K^+$.
The deviation of the value $3$ from the heavy quark limit value $1$ indicates significant heavy quark symmetry breaking effect.
 Thus the suppression from the charge-parity is not sizeable.
In conclusion, the production rate of the $\Zc$ in $B$ decays
is dominantly suppressed by the small value of the isospin triplet production amplitude $B_1$.
  
On the contrary, in $e^+e^-$ annihilation, since the $\Zc$ and the $\X$ are produced 
together with $\pi$ and $\gamma$ emissions, respectively, 
the C-parity will not suppress any of them. At the same time, since virtual photons do 
not have fixed isospin,
the isospin factors for both isospin triplet and isospin singlet are the same.  
The only suppression happens for the production of the $\X$ in $e^+e^-$ annihilation 
stemming from the additional factor of the fine-structure constant. 

In summary, we have analyzed the isospin amplitudes of the $B\to D^*\bar{D}K$ 
process and fitted the presently available $D^{*0}\bar{D}^0$ invariant mass distributions.
The results indicate that the production of the isospin triplet $D^*\bar{D}$ state is highly 
suppressed due to the small value of $b_1$. 
In addition, the $\Zc$ will be further suppressed because of its negative charge-parity. 
These two reasons for the first time lead to a concise explanation 
why the $\Zc$ is absent in $B$ decays  within the hadronic molecular picture.
Stated differently, the absence of the $\Zc$ in $B$ decays clearly points towards 
its $D^*\bar{D}$ molecular nature.  
A detailed scan of the $D^*\bar{D}$ invariant mass distributions of the six processes 
$B^0\to D^{*0} {D}^{-}K^+$, $B^0\to D^{*+} {D}^{-}K^0$, $B^0\to D^{*0} \bar{D}^{0}K^0$, 
$B^+\to D^{*+} \bar{D}^{0}K^0$, $B^+\to D^{*0} \bar{D}^{0}K^+$, $B^+\to D^{*+} {D}^{-}K^+$ with high accuracy 
(near threshold), especially for the $B^0\to D^{*0} {D}^{-}K^+$ and $B^+\to D^{*+} \bar{D}^{0}K^0$ reactions
 where only the isospin triplet production amplitude contributes,
 will help to reduce the uncertainty of the conclusion, 
 such as the uncertainty of the ratios between the isospin triplet amplitudes and the isospin singlet ones.

\medskip
Q.W. is grateful to Yu-Ming Wang for useful discussions and comments. 
This work is
supported in part by the DFG and the NSFC through funds provided to
the Sino-German CRC 110 ``Symmetries and the Emergence of Structure
in QCD''. The work of U.G.M. was also supported by the Chinese Academy 
of Sciences (CAS) President's International Fellowship Initiative (PIFI) 
(Grant No. 2017VMA0025). The work of Z.Y. was also supported
 by CAS Pioneer Hundred Talents Program.

\end{document}